\documentclass[%
 aps,
 prl,
preprint,%
superscriptaddress%
]{revtex4-1}

\usepackage{graphicx}
\usepackage{dcolumn}
\usepackage{bm}
\usepackage{color}

\usepackage{acronym}

\newacro{AUG}[AUG]{ASDEX Upgrade}
\newacro{CXRS}[CXRS]{charge exchange recombination spectroscopy}
\newacro{BES}[BES]{beam emission spectroscopy}
\newacro{betaN}[$\beta\rm{_N}$]{normalized beta}
\newacro{bt}[$|B\rm{_T}|$]{toroidal field}
\newacro{pe}[$p\rm{_e}$]{electron pressure}
\newacro{DBP}[$D_{\omega}$]{distribution of the observed intensity}
\newacro{DW}[DW]{drift-wave}

\newacro{ne}[$n\rm{_e}$]{electron density}
\newacro{overne}[$\overline{n\rm{_e}}$]{line averaged $n\rm{_e}$}

\newacro{Te}[$T\rm{_e}$]{electron temperature}
\newacro{fIF}[$f\rm{_{IF}}$]{IF bandwidth}
\newacro{Trad}[$T\rm{_{rad}}$]{radiation temperature}
\newacro{Ti}[$T\rm{_i}$]{ion temperature}
\newacro{DCN}[DCN]{deuterium cyanide}
\newacro{dphi}[$\Delta \varphi\rm{_{UL}}$]{differential phase angle}
\newacro{ECE}[ECE]{electron cyclotron emission}
\newacro{ECFM}[ECFM]{electron cyclotron forward modeling}
\newacro{ECEI}[ECE-I]{ECE-imaging}
\newacro{IF}[IF]{intermediate frequency}
\newacro{inte}[$I_{imp}$]{measured intensity from impurity}

\newacro{FEM}[FEM]{finite element calculations}
\newacro{KBM}[KBM]{kinetic ballooning mode}

\newacro{ICRH}[ICRH]{ion cyclotron resonance heating}
\newacro{ECR}[ECR]{electron cyclotron resonance}
\newacro{P_ECRH}[$P\rm{_{ECRH}}$]{ECRH power}
\newacro{OH}[OH]{ohmic heating}
\newacro{P_OH}[$P\rm{_{OH}}$]{ohmic heating power}
\newacro{PSL}[PSL]{passive stabilization loop}
\newacro{P_NET}[$P\rm{_{NET}}$]{applied net heating power}
\newacro{LIB}[LIB]{lithium beam}
\newacro{LSQ}[LSQ]{least square}
\newacro{LFS}[LFS]{low field side}
\newacro{HFS}[HFS]{high field side}
\newacro{RFA}[RFA]{resonant field amplification}
\newacro{REF-O}[REF-O]{O-mode reflectometer}
\newacro{REF-X}[REF-X]{X-mode reflectometer}

\newacro{LCFS}[LCFS]{last closed flux surface}
\newacro{ELM}[ELM]{edge localized mode}
\newacro{H-mode}[H-mode]{high confinement mode}
\newacro{L-mode}[L-mode]{low confinement mode}

\newacro{WMHD}[$W\rm{_{MHD}}$]{plasma energy}
\newacro{nustar}[$\nu^{\star}$]{collisionality}

\newacro{MP}[MP]{magnetic perturbation}
\newacro{IDA}[IDA]{integrated data analysis}
\newacro{L-mode}[L-mode]{low confinement mode}
\newacro{NBI}[NBI]{neutral beam injection}
\newacro{SOL}[SOL]{scrape off layer}
\newacro{LOS}[LOS]{lines of sight}
\newacro{MHD}[MHD]{magnetohydrodynamic}
\newacro{PBM}[PBM]{peeling-ballooning mode}

\newacro{TS}[TS]{Thomson scattering}

\begin{document}

\title{Field-line localized destabilization of ballooning modes in 3D tokamaks}

\author{M.~Willensdorfer}
\email{matthias.willensdorfer@ipp.mpg.de}
\affiliation{Max Planck Institute for Plasma Physics, 85748 Garching, Germany}
   \author{T.B. Cote}
   \email{tcote@wisc.edu}
\affiliation{University of Wisconsin-Madison, Madison, Wisconsin 53706, USA}
   \author{C.C. Hegna}
   \affiliation{University of Wisconsin-Madison, Madison, Wisconsin 53706, USA}

   \author{W. Suttrop}
   \affiliation{Max Planck Institute for Plasma Physics, 85748 Garching, Germany}

   \author{H. Zohm}
\affiliation{Max Planck Institute for Plasma Physics, 85748 Garching, Germany}

  \author{M. Dunne}
  \affiliation{Max Planck Institute for Plasma Physics, 85748 Garching, Germany}
  
  \author{E. Strumberger}
  \affiliation{Max Planck Institute for Plasma Physics, 85748 Garching, Germany}

 \author{G.~Birkenmeier}
 \affiliation{Max Planck Institute for Plasma Physics, 85748 Garching, Germany}
\affiliation{Physik-Department E28, Technische Universit\"at M\"unchen, 85748 Garching, Germany}

 \author{S.S.~Denk}
 \affiliation{Max Planck Institute for Plasma Physics, 85748 Garching, Germany}
\affiliation{Physik-Department E28, Technische Universit\"at M\"unchen, 85748 Garching, Germany}

  \author{F.~Mink}
  \affiliation{Max Planck Institute for Plasma Physics, 85748 Garching, Germany}

 \author{B. Vanovac}
 \affiliation{FOM-Institute DIFFER, Dutch Institute for Fundamental Energy Research }
 
  \author{L.C. Luhmann}
 \affiliation{University of California at Davis, Davis, CA 95616, USA }

\author{the ASDEX Upgrade Team}
 \affiliation{Max Planck Institute for Plasma Physics, 85748 Garching, Germany}

   
 \begin{abstract}
Field-line localized ballooning modes 
have been observed at the edge of high confinement mode plasmas in ASDEX Upgrade with rotating 3D perturbations induced by an externally applied $n=2$ error field and during a moderate level of \acl{ELM}-mitigation.
The observed ballooning modes are localized to the field-lines which experience one of the two zero-crossings of the radial flux surface displacement during one rotation period.
The localization of the ballooning modes agrees very well with the localization of the largest growth rates from infinite-n ideal ballooning stability calculations using a realistic 3D ideal \acl{MHD} equilibrium. This analysis predicts a lower stability with respect to the axisymmetric case.
The primary mechanism for the local lower stability is the 3D distortion of the local magnetic shear.
 \end{abstract}


 \maketitle
 
\acresetall

 \textit{Introduction -} In order to mitigate the possible harmful heat load from \acp{ELM} in the \ac{H-mode} in future fusion devices, it is necessary to suppress or to mitigate the \acp{ELM}.   The application of  \acp{MP} enables mitigation of \acp{ELM} and under certain circumstances even suppression. 
 This method has the side-effect of a reduced plasma density at low \ac{nustar}, and reduced pedestal plasma pressure, the so-called density 'pump-out'~\cite{Evans:2004}.

  In recent years, there is growing evidence that stable ideal kink modes can amplify the externally applied \ac{MP}-field~\cite{Liu:2011,Paz-Soldan:2015}, which plays a key role in  \ac{ELM}  mitigation~\cite{Kirk:2015} and  \ac{ELM} suppression at low \ac{nustar}~\cite{Nazikian:2015, Suttrop:2017}.  
   These \ac{MHD} modes are driven by the edge pressure gradient  and/or the associated bootstrap current. 
Comparative studies in combination with \ac{MHD} modeling indicate that the highest \ac{ELM} frequency,  the strongest density 'pump-out' and  the strongest accompanying reduction in the edge pressure gradient are correlated with the coupling of the stable ideal kink modes to resonant components~\cite{Paz-Soldan:2016, Liu:2016, Orain:2016}. The resulting 3D boundary distortion can be many times larger than expected solely from the \ac{MP} of the vacuum field~\cite{Chapman:2014a,Moyer:2012}. 
It has been argued that the \acp{MP} modify the \ac{PBM} stability by a change of edge  bootstrap current due to the 'pump-out'~\cite{Chapman:2013} or equilibrium currents around rational surfaces~\cite{Ham:2015}. To approximate changes in the finite-$n$ \ac{PBM} stability~\cite{Chapman:2013}, the local infinite-$n$ ideal MHD ballooning theory is often used as an estimate. 

The infinite-$n$ ballooning theory in 3D \ac{MHD} geometry has been extensively studied for stellerator configurations~e.g.~Ref~\cite{Talmadge:1996, Hegna:1998}.
First theoretical attempts to apply  it to a 3D tokamak geometry induced by external \acp{MP} were done in Ref.~\cite{Bird:2013}. It was proposed 
that the 3D modulations of the local magnetic shear associated with the presence of near-resonant Pfirsch-Schl\"uter currents peturb the ballooning stability boundary. This can lead to enhanced growth rates  on local field-lines. 

In this Letter, we present novel first measurements, which demonstrate field-line localized destabilization of ideal \ac{MHD}  ballooning modes in the presence of an edge perturbed 3D \ac{MHD} tokamak equilibrium. The measured localization of the ballooning modes in the 3D geometry combined with infinite-$n$ ballooning calculations supports the proposed importance of the changes in the local magnetic shear.

\begin{figure}[ht]
\includegraphics[width=0.5\textwidth]{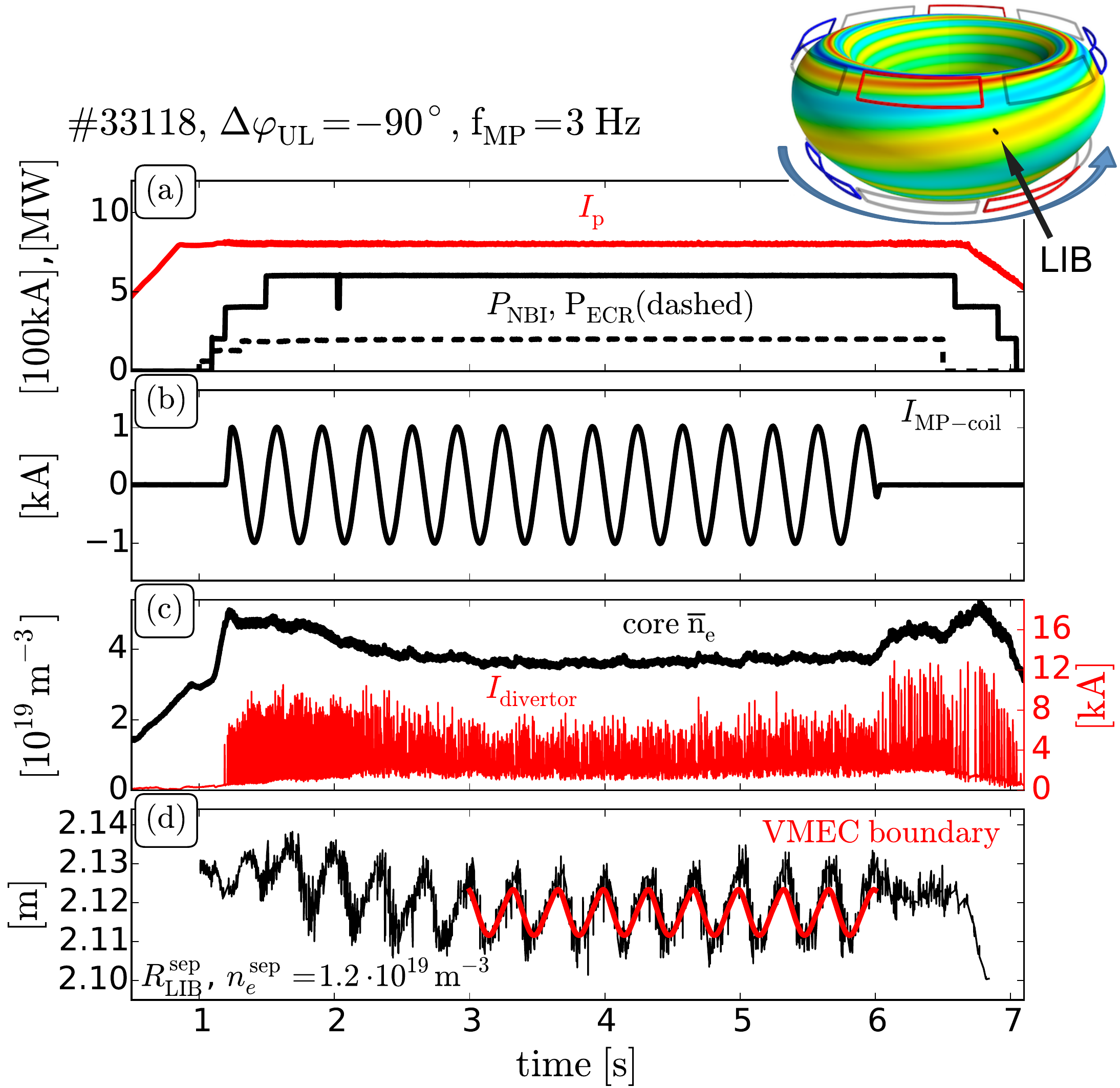} 
\caption{ 
Time traces from a discharge with a rigidly rotating \ac{MP}-field resulting in a rotation of the $\xi\rm{_r}$. Inset shows corresponding $\xi\rm{_r}$ from VMEC (red outwards, green zero, blue inwards at the boundary) and coil configuration at $4\ \rm{s}$. (a) plasma current, external \ac{ECR} and \ac{NBI} heating power, (b) supply current of one coil, (c) core \ac{overne} and divertor current and (d) separatrix movements along the \ac{LIB} from measured pre-\ac{ELM} \ac{ne} profiles and VMEC boundary. Good agreement is found.
}
\label{Overview3D}
\end{figure}

  \textit{Experimental setup -} The presented example is from a series of ASDEX Upgrade experiments to measure the displacement by combining rigidly rotating \ac{MP}-fields with the toroidal mode number $n=2$ and toroidally localized diagnostics~\cite{Willensdorfer:2017}. In this work, we primarily use  \ac{ECE} diagnostics, which deliver the \ac{Te} around the \ac{LFS} midplane from the measured \ac{Trad}~\cite{Willensdorfer:2016}.
  To have the best coverage at the edge for profile-\ac{ECE} and \ac{ECEI}~\cite{Classen:2014} measurements, we set a \ac{bt} of around $2.5\ \rm{T}$.
  Further global parameters are a plasma current $I_P$ of $800\ \rm{kA}$, edge safety factor $q_{95}$ of $5.3$ and a core \ac{ne} of $4.3\;\rm{10^{19}m^{-3}}$. The applied \ac{NBI} and \ac{ECR} heating are approximately 6 and $2\;\rm{MW}$, respectively (see Fig.~\ref{Overview3D}(a)). A slow ($f\rm{_{MP}}=3\ \rm{Hz}$) rotating \ac{MP}-field is applied. The \ac{MP}-coil supply current shown in Fig.~\ref{Overview3D}(b) illustrates the timing.
  The rotation is in the positive toroidal direction using a fixed \ac{dphi} of around $-90^\circ$ between the \ac{MP}-field from the upper and lower coil set (see cartoon in right top corner of Fig.~\ref{Overview3D}). Because of 8 coils in each row and a low intrinsic error-field, the dominant $n=2$ spectrum varies little during the rotation as seen by small core \ac{ne} perturbations ($<5\%$). 

The applied \ac{MP}-field configuration ($\ac{dphi}\approx-90^\circ$) is optimal to excite stable ideal kink modes at the edge~\cite{Liu:2016}, which leads to a moderate level of density 'pump-out' and a reduction of the \ac{ELM} size. This is best seen in Fig.~\ref{Overview3D}(c) at $6\;\rm{s}$ during the switch-off of the \ac{MP}-field.

The rigidly rotating \ac{MP}-field causes a rotation of the radial displacement ($\xi_r$, normal to axisymmetric flux surface) at the boundary, which is seen as a modulation in edge profile diagnostics. To track $\xi\rm{_r}$ at the boundary, we use the \ac{ne} profiles from the \ac{LIB} diagnostic~\cite{Willensdorfer:2014}  assuming a constant separatrix density (Fig.~\ref{Overview3D}(d)).  

To model $\xi_r$, we employ the 3D ideal \ac{MHD} equilibrium code VMEC~\cite{Hirshman:1983}. If no strong resistive \ac{MHD} modes are active, VMEC is able to predict $\xi_r$ at the edge~\cite{Chapman:2014a, Willensdorfer:2017}. 
     For details about its setup for ASDEX Upgrade discharges and the accuracy of VMEC in the presence of rational surfaces, we refer to Ref.~\cite{Strumberger:2014, Willensdorfer:2016, Willensdorfer:2017} and Ref.~\cite{Loizu:2016, Lazerson:2016}, respectively. 
For the given configuration VMEC predicts a maximum $\xi_r$ of $\pm11\;\rm{mm}$ at the plasma top and  $\pm6\;\rm{mm}$ around the \ac{LFS} midplane (visualized in the inset of Fig.~\ref{Overview3D}). To underline the agreement of the calculated $\xi_r$ with experiments, the predicted boundary corrugation from the 3D VMEC equilibrium along the \ac{LOS} of the \ac{LIB} is added in Fig.~\ref{Overview3D}(d). This equilibrium is calculated once at $4\;\rm{s}$ and the corrugation is then mapped onto the time base according to the phase of the \ac{MP}-field rotation~\cite{Willensdorfer:2016}. Good agreement is found and the amplitudes differ by not more than $1.3\;\rm{mm}$.

\begin{figure}[ht]
\includegraphics[width=0.5\textwidth]{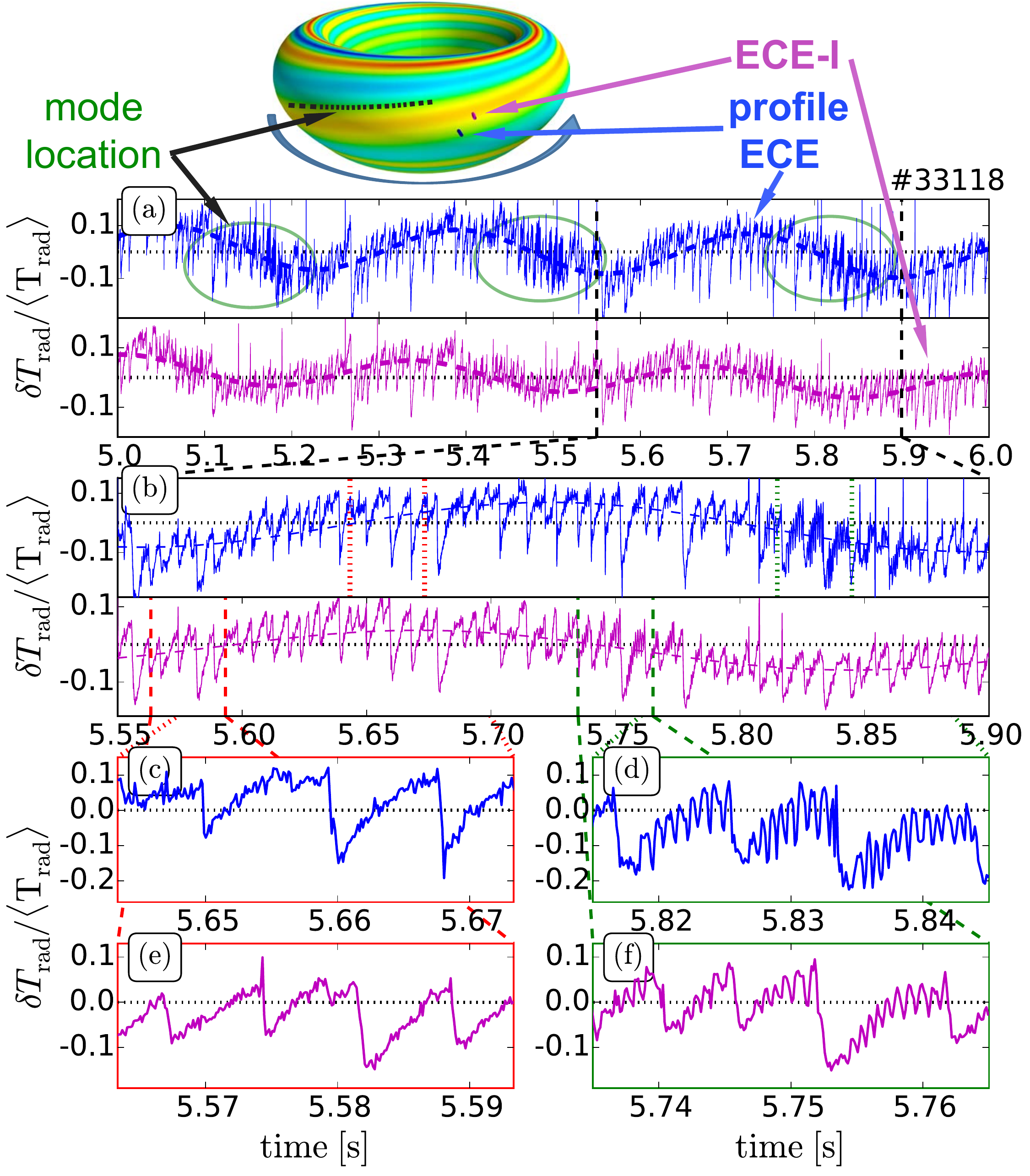} 
\caption{Time traces from profile \ac{ECE} (blue) and \ac{ECEI} (purple) channels. Measuring principle and \ac{LOS} positions are indicated at the top.
(a) $3\;\rm{Hz}$ modulation due to the rotating $\xi_r$, (b) a \ac{MHD} mode is clearly seen when one specific $\xi\rm{_r}\approx0$ (horizontal dotted line) passes the diagnostics. The mode appears in-between \ac{ELM} crashes and at only one $\xi_r\approx0$ (d, f), whereas at the other times not e.g.~(c, e). Please note the different time ranges between (c-f). }
\label{ECEtraces}
\end{figure}

 \textit{Observations -} During rigid rotation experiments, we observe an instability occurring only at certain toroidal phases.  Figure~\ref{ECEtraces} shows time traces of the measured relative amplitude ($\delta \ac{Trad}/ \langle \ac{Trad}\rangle $ with  $\delta \ac{Trad}=\ac{Trad} - \langle\ac{Trad} \rangle$), where $\langle \ac{Trad}\rangle$ is averaged over the 3 periods, 
 from one profile \ac{ECE} (blue) and one \ac{ECEI} (magenta) channel probing the steep gradient region near the pedestal top. These channels are continuously optically thick, so we can assume $\ac{Te}\approx\ac{Trad}$.
 Three rotation periods and  \ac{ELM} crashes are seen Fig.~\ref{ECEtraces}(a). 
 Fits of sinusoidal functions (dashed colored lines) using only pre-\ac{ELM} data points emphasize the modulation from the rotating $\xi\rm{_r}$.
 Because of the poloidally as well as toroidally separated measurement  positions and the alignment of  $\xi\rm{_r}$ (see cartoon in Fig.~\ref{ECEtraces}), the two diagnostics exhibit a relative phase shift of around $\pi/2$. 
 
 High frequency modes ($f\approx1\;\rm{kHz}$) appear in the diagnostics, when field-lines with specific 3D geometry pass their \ac{LOS} (green circles). To enhance their visibility, Fig.~\ref{ECEtraces}(b) magnifies one period. It is clearly seen that these modes develop at different times in the profile \ac{ECE} (Fig.~\ref{ECEtraces}(d)) and \ac{ECEI} channel (Fig.~\ref{ECEtraces}(f)), but at the same toroidal phase with respect to the $3\;\rm{Hz}$ modulation. They appear in-between \acp{ELM} and during one period they only occur  once when the modulation crosses zero (dotted line) from positive to negative values (Fig.~\ref{ECEtraces}(d,f)). At other times, such pronounced oscillations are not seen e.g.~Fig.~\ref{ECEtraces}(c,e).
 The amplitude of mode appears to be smaller in the \ac{ECEI} in comparison to the profile \ac{ECE} (Fig.~\ref{ECEtraces}(d,f)). This is simply because of the twice as large observation volume of the \ac{ECEI} resulting in a smearing of the amplitude.

\begin{figure}[ht]
\includegraphics[width=0.5\textwidth]{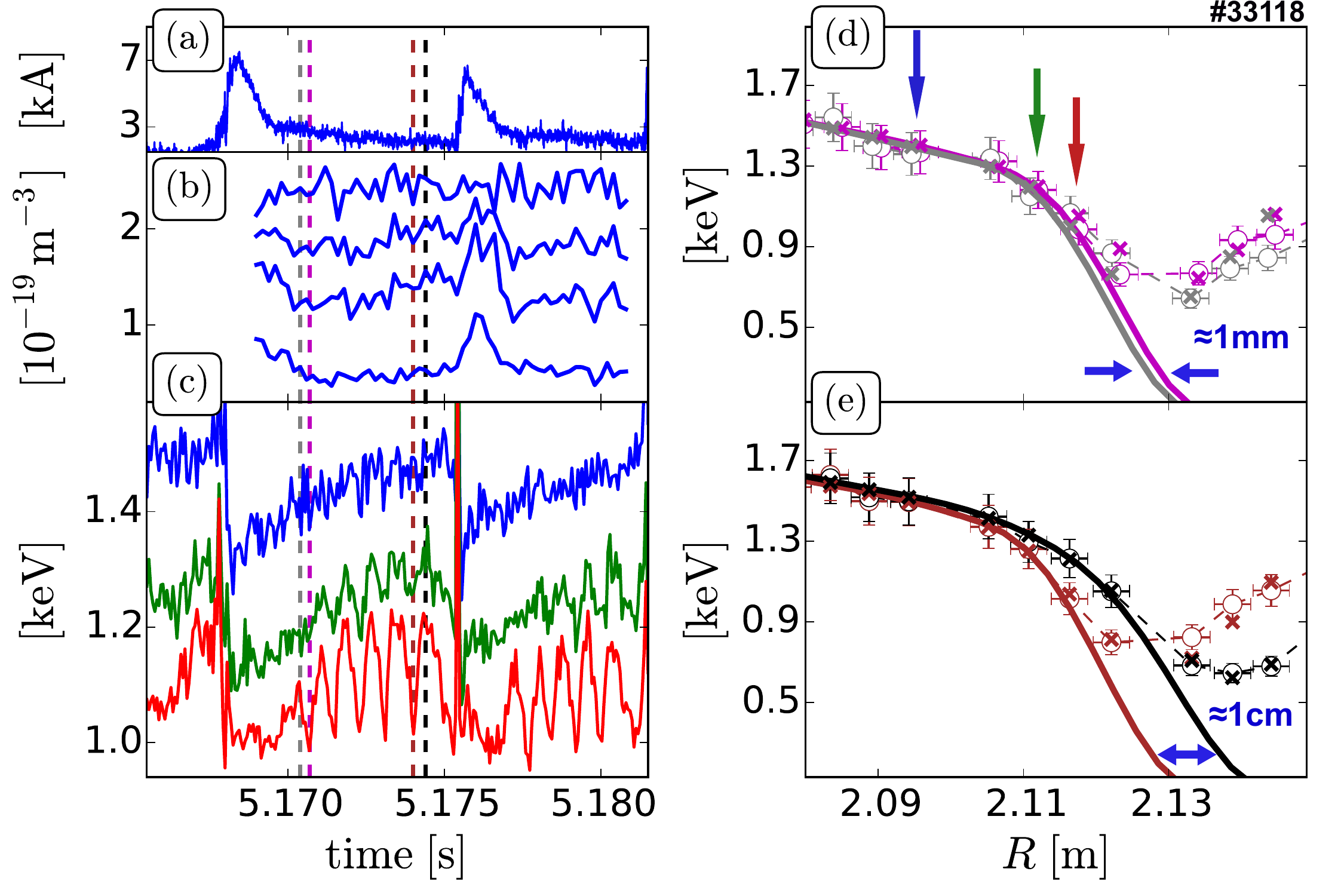} 
\caption{(a) divertor currents (b) edge \ac{ne} from \ac{LIB} and (c) \ac{Trad} in gradient region. (d, e) \ac{Te} profiles (solid lines) and the corresponding modeled (x) and measured \ac{Trad} (o$--$). The timing of the profiles are indicated by colored vertical lines in (a-c). Arrows in (d) indicate \ac{ECE} channels used in (c). Distortion amplitude of the $1\;\rm{kHz}$ mode increases during the pedestal recovery.}
\label{ECEprofile}
\end{figure}

To demonstrate the properties of an ideal \ac{MHD} instability, Fig.~\ref{ECEprofile} shows  \ac{Te} profiles during the \ac{ELM} recovery determined from forward-modeling of the measured  \ac{Trad} profiles~\cite{Rathgeber:2013}.   
After the  \ac{ELM} crash, the edge pressure profile recovers as indicated by a slight steepening of \ac{ne} in the edge gradient region (Fig.~\ref{ECEprofile}(b)) and an increase of \ac{Te} at the pedestal top  (blue in Fig.~\ref{ECEprofile}(c)).
Simultaneously, the distortion amplitude of the high frequency mode increases from less than $1\;\rm{mm}$ to around $1\;\rm{cm}$ (Fig.~\ref{ECEprofile}(d,c)), which is suggestive for a pressure gradient driven instability. 
Clear radial displacements in the pedestal are observed.
There is no indication of a magnetic island such as profile flattening at a rational surface or the \ac{Te}  perturbations being in anti-phase on both sides of a rational surface.

The high frequency modes are also seen in other profile diagnostics around the \ac{LFS} midplane like \ac{LIB}. They are also measured at the \ac{LFS} by soft X-ray, diode bolometers and weakly in $B\rm{_{\Theta}}$ probes, but these diagnostics do not detect these modes at the \ac{HFS}.
This  gives us confidence that we are dealing with ideal \ac{MHD} ballooning modes. 
Moreover, the mode rotates toroidally into the opposite direction of the \ac{MP}-field rotation, so clockwise and poloidally into the electron diamagnetic direction as expected from ballooning modes.

These ballooning modes  are also observed in other discharges from this experimental series and also in similar experiments at $\ac{bt}\approx2.0\ \rm{T}$ with dominant \acl{ICRH}. Their appearance seems to be correlated with a sufficiently large $\xi_r$ around \ac{LFS} midplane and \ac{betaN}. Although these experiments have different applied poloidal mode spectra set by \ac{dphi} \cite{Willensdorfer:2017}, the modes always occur around the same field-lines in the 3D geometry, which experience the zero-crossing $\xi_r$ from positive to negative values when the \ac{MP}-field is rotated in positive toroidal direction.
Moreover, the $2.0\ \rm{T}$ experiments in which the \ac{MP}-field amplitude, normalized to the background magnetic field, is larger, show
coupling of the modes to the external \ac{MP}-field resulting in  a vanishing  phase velocity relative to the $n=2$ motion.

In summary, we observe rotating ballooning modes, which are only destabilized around a certain field-line in the 3D flux surface geometry.
 These ballooning modes do not appear at the maximum $\xi\rm{_r}$, where the local flux expansion causes  the largest curvature and the smallest peak pressure gradients in real space $\partial p/\partial r$.  They are also not at the minimum $\xi\rm{_r}$, where the  local flux compression leads to the largest pressure gradients. This is different to the observed ballooning modes in the presence of internal kink modes in the core. In these cases the largest gradients come with the largest displacements, which then locally  destabilizes  the ballooning modes in the core~\cite{Park:1995}.
In the presence of perturbed flux surfaces from externally applied \acp{MP}, the modes are dominantly destabilized, where $\xi\rm{_r}$ is zero as predicted by Ref.~\cite{Bird:2013}.
Even more interestingly, they only appear at one specific zero-crossing although there are two in each period. The reason for this will be elucidated in the following. 
 
 \begin{figure}[ht]
\includegraphics[width=0.5\textwidth]{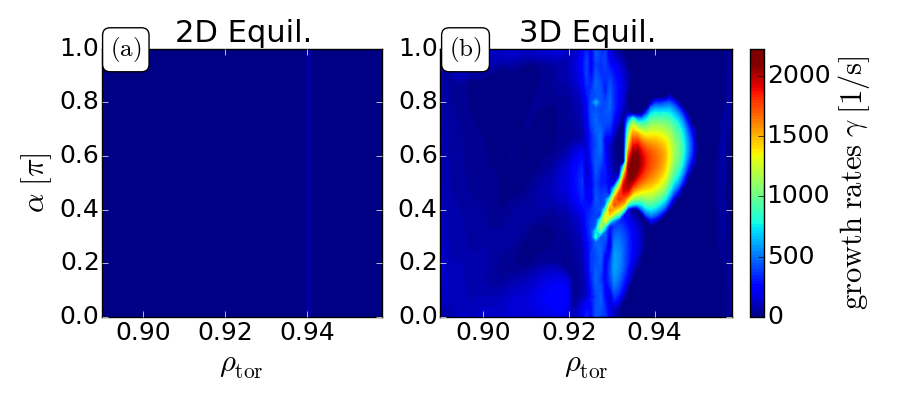} 
\caption{Growth rates $\gamma$ on the field-line label $\alpha$ versus $\rho\rm{_{tor}}$ for (a) axisymmetric 2D and (b) perturbed 3D equilibrium. The 3D case has locally enhanced $\gamma$ and is most unstable around $\alpha=0.6\pi$.}
\label{growthrates}
\end{figure}
\pagebreak
  \textit{Stability analysis -}    Motivated by the experimental observations, we extended the stability analysis from Ref.~\cite{Bird:2013} using the 3D equilibrium from VMEC  introduced previously and its $n=0$ solution for the 2D case.  The addition of 3D magnetic perturbations causes the local plasma stability to vary on different magnetic field-lines in the steep gradient region of the pressure profile.  
  Applying the ideal ballooning theory allows us to solve for the high-$n$, local plasma stability on individual field-lines. 
The calculated growth rates $\gamma$ (color scaling) in the pedestal 
are shown in Fig.~\ref{growthrates} as a function of the field-line label $\alpha=q\Theta^\star-\phi$ with the straight field coordinates ($\phi,\Theta^\star$)~\cite{Hirshman:1983} versus the normalized toroidal flux ($\rho\rm{_{tor}}$) for one $n=2$ period. In comparison to the axisymmetric $n=0$ case (Fig.~\ref{growthrates}(a)), the growth rates of the perturbed equilibrium are enhanced (Fig.~\ref{growthrates}(b)), which indicates a lower stability in the 3D case with respect to the axisymmetric case. The growth rates are largest around the field-line $\alpha=0.6\pi$.

\begin{figure}[ht]
\includegraphics[width=0.5\textwidth]{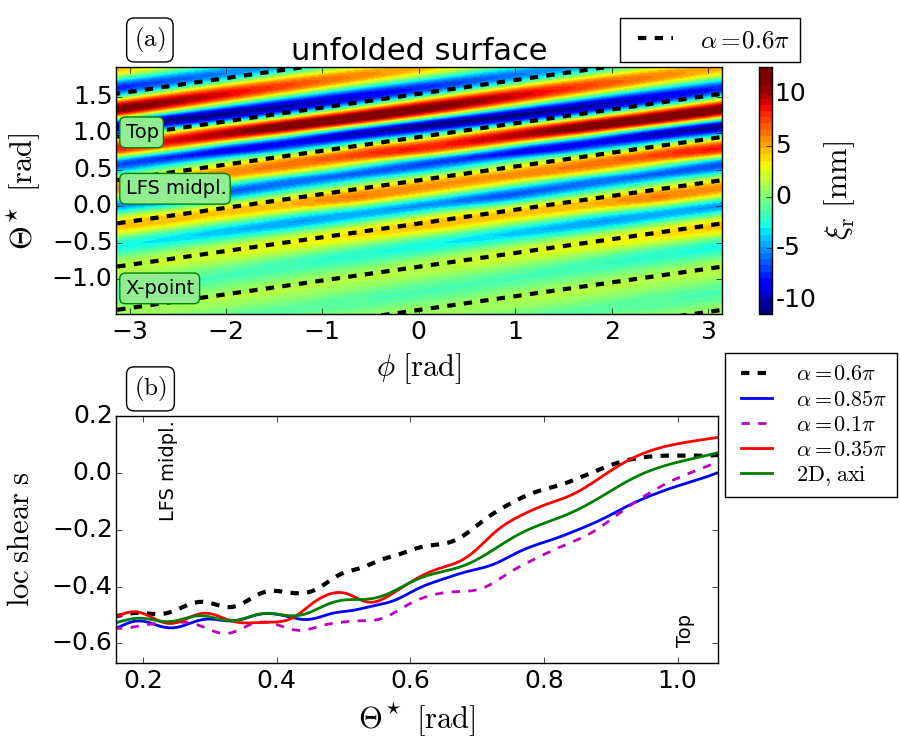} 
\caption{ (a) Unfolded surface at $\rho\rm{_{tor}}\approx0.935$ showing the radial displacement $\xi_r$.  Positive values are pointing outwards. The dashed line indicates the most unstable field-line $\alpha=0.6\pi$. (b) Local magnetic shear $s$ versus $\Theta^\star$ for various field-lines. The most unstable field-line has the lowest magnitude of $s$.}
\label{unfolded}
\end{figure}

To identify the position of the most unstable field-line $\alpha=0.6\pi$ in the 3D geometry, Fig.~\ref{unfolded}(a) shows $\xi\rm{_r}$ on  the unfolded axisymmetric flux surface and this field-line. It is located at the same zero-crossing as the measured localization of  the ballooning mode in Fig.~\ref{ECEtraces}.
 To elucidate the reason for this specific field-line to be unstable, Figure~\ref{unfolded}(b) shows the corresponding local magnetic shear (defined $s = \hat{b} \times \hat{n} \cdot \nabla \times (\hat{b} \times \hat{n})$ where $\hat{b}$ and $\hat{n}$ are the unit magnetic field and normal components to the magnetic surface, respectively~\cite{Hegna:2000}) versus $\Theta^\star$ for several field-lines. $s$ enters the ballooning equation~\cite{Bird:2013} and stabilizes ballooning modes against field-line bending. Thus, the most unstable field-line exhibits the lowest absolute values of $s$ in the region of the negative normal and positive geodesic curvature (region between \ac{LFS} midplane and plasma top, Fig.~\ref{unfolded}(b)). Although the changes in $s$ are small, they are enough to significantly vary  the growth rates of the ballooning modes as seen in Fig.~\ref{growthrates}.

The change in local magnetic shear is related to changes in the local parallel current profile and geometric shape of the magnetic surface as described by the identity $s = \mu_0 J_{\parallel} / B - 2 \tau_n$, where the normal torsion is defined by $\tau_n = - \hat{n} \cdot (\hat{b} \cdot \nabla ) \hat{b} \times \hat{n}$~\cite{Hegna:2000}. Near rational surfaces in 3D MHD equilibrium, Pfirsch-Schl\"uter currents become resonant and hence can produce large distortions in local shear~\cite{Bird:2013}. However, analysis of the equilibrium presented here show the 3D distortion produced changes the normal torsion that corresponds to minimize the region of local shear in the vicinity of the field line $\alpha = 0.6 \pi$~\cite{Cote:2017}. The 3D distortions, however, do not produce appreciable changes to field-line curvature indicating that the dominant reason for ballooning destabilization is the reduction of stabilizing field-line bending.

 \textit{Discussion -} The presented measurements combined with 3D ideal \ac{MHD} modeling evidence the importance of the additional 3D geometry on the  infinite-n ballooning stability. 
The localization of the ballooning modes identifies the variation of the local magnetic shear as the dominant mechanism in the additional destabilization.
Our observation and analysis suggest that the reduced edge pedestal pressure in H-mode plasmas with non-axisymmetric MP is due to a modification of the edge stability boundary introduced by the 3D distortion of the local magnetic shear. Furthermore, it could also deliver reasonable explanations for enhanced transport, the density 'pump-out', since the local magnetic shear  can also  influence  further curvature driven instabilities like \acp{KBM} and  \acp{DW}~\cite{Kendl:1999,Birkenmeier:2013}. 
Nevertheless, dedicated studies using 3D stability calculations of e.g.~\ac{PBM}~\cite{Weyens:2017}, \ac{KBM}~\cite{Holod:2017}, \ac{DW}~\cite{Kendl:1999} are needed to pin down their role and the role of the 3D equilibrium in the \ac{ELM} mitigation/suppression. 
Finally, we would like to point out that the described mechanism is solely based on single fluid ideal \ac{MHD} physics and does not invoke any mode penetration~\cite{Nazikian:2015} or  ergodization~\cite{Becoulet:2014}. 

\textit{Acknowledgments -}
This research was supported by US Department of Energy under grant no.~DE-FG02-86ER53218.
M.~W.~has received funding from the Euratom training programme 2014-2018 under grant agreement No 633053. The views and opinions expressed herein do not necessarily reflect those of the European Commission.

\bibliographystyle{unsrt}
\bibliography{ballooning}

\end{document}